\documentclass[12pt]{iopart}
\newcommand{\xv}{\mbox{\boldmath$x$}}
\usepackage{graphicx}
\usepackage{amssymb}

\begin{document}
\title[Exact location of the multicritical point: A conjecture]{Exact
location of the multicritical point for finite-dimensional spin glasses:
 A conjecture}
\author{Koujin Takeda, Tomohiro Sasamoto and Hidetoshi Nishimori}
\address{Department of Physics, Tokyo Institute of Technology,
Oh-okayama, Meguro-ku, Tokyo 152-8551, Japan}

\begin{abstract}
We present a conjecture on the exact location of the multicritical point in
the phase diagram of spin glass models in finite dimensions.
By generalizing our previous work, we combine duality and gauge symmetry
for replicated random systems to derive formulas which make it possible
to understand all the relevant available numerical results in a unified way.
The method applies to non-self-dual lattices as well as to self dual cases,
in the former case of which we derive a relation for a pair of values 
of multicritical points for mutually dual lattices.
The examples include the $\pm J$ and Gaussian Ising spin glasses on the square,
hexagonal and triangular lattices, the Potts and $Z_q$ models with chiral
randomness on these lattices, and the three-dimensional $\pm J$ Ising spin
glass and the random plaquette gauge model.
\end{abstract}


\section{Introduction}

Properties of finite-dimensional spin glasses are still under debate
although the problem is essentially settled for the mean-field model
\cite{SG-general}.
Outstanding problems for finite-dimensional spin glasses include the
existence or absence of spin glass phase and whether or not the mean-field
picture of the spin glass phase applies.
Another interesting, but less extensively studied, issue is the structure
of the phase diagram, in particular where precisely the multicritical point
is located and what the values are for the critical exponents characterizing the system
behaviour at and away the multicritical point.
The present paper discusses this problem of the location of the multicritical
point for finite-dimensional spin glass models by analytical methods.

A number of numerical investigations on this problem exist for various lattices.
However, it has been quite difficult to derive analytical results for regular
finite-dimensional lattices until a few years ago when we succeeded in devising
a method to predict the exact locations of the multicritical points for the
square lattice Ising and Potts models and four-dimensional random plaquette
gauge model using duality, gauge symmetry and the replica method
\cite{NN,MNN,TN}. In the present paper
we generalize this theoretical framework so that it is applicable to
a pair of mutually-dual lattices, for which our theory relates the
pair of values of multicritical points of the two lattices.

The logic of our theory includes a step which is yet to be justified rigorously,
and hence the status of our result is a conjecture at this moment.
Nevertheless, our theory enables us to understand all the relevant available numerical data
for the multicritical points derived independently by a number of authors.
Also our theory satisfies necessary conditions which the
exact solution should satisfy.

This paper is organized as follows. In the next section, 
we recall the basic formulation which was developed in our previous studies
in order to fix the notation and set the stage for further developments in the
following sections.
In section 3, a generalization of the theory to non-self-dual $Z_2$ (Ising)
models with randomness is discussed.  This argument is followed by section 4,
in which we further generalize the arguments to $Z_q$ systems with chiral randomness.
The final section is devoted to conclusion and discussions.


\section{Self-dual $Z_2$ models}

In this section we briefly review the duality arguments 
in \cite{MNN,TN} applied to the two-dimensional ($2d$) $\pm J$
random bond Ising model on the square 
lattice and the $4d$ random $Z_2$ lattice gauge model, by which we fix the
notation and set a stage for generalizations in the following sections.
After recalling the duality arguments for non-random systems, 
we apply the idea to random systems.

\subsection{Duality of non-random models}

 We first elucidate the duality of a
 non-random $Z_2$ (Ising) system \cite{Wegner}.
 Let us prepare a $d$-dimensional lattice and assign $Z_{2}$
 spins on $r-1$ dimensional elements $\xv$ on
 the lattice, which we denote by $S_{\xv}$.
 We consider a model on the lattice whose Hamiltonian is given by
\begin{equation}
 \label{generalspin}
 H = -J \sum_{C} \prod_{\xv \in \partial C} S_{\xv}, 
 \end{equation} 
where $C$ is the $r$-dimensional element on the lattice and 
 $\partial C$ is its boundary of dimension $r-1$ \cite{Wegner}. 
 Let $u_{\pm 1}$ denote the Boltzmann factor for an element $C$,
 \begin{equation}
 u_{\pm 1}(K) \equiv e^{\pm K}, \hspace{1cm} {\rm for}~
 \prod_{\xv \in \partial C} S_{\xv} = \pm 1,
 \end{equation}
where $K \equiv \beta J$.
 For the case of the usual Ising model $(r=1)$, $u_1(K)$ is the bond (edge)
 Boltzmann factor
 for parallel spins at both ends and $u_{-1}(K)$ is for anti-parallel spins.
 Then the partition function ${\cal Z}$ is a function of $u_{\pm 1}$;
 ${\cal Z} = {\cal Z}\{u_1(K),u_{-1}(K)\}$.

 The dual model is defined on the dual lattice (the definition of
 which is given in \cite{Wegner}).
 The dual Boltzmann factor for the dual element $C^*$ is 
 defined by the two-component Fourier transformation of 
 $u_{\pm 1}$ \cite{WuWang},
 \begin{equation}
 {u^{*}_{\pm 1}}(K) \equiv \frac{u_1(K) \pm u_{-1}(K)}{\sqrt{2}}
  = \frac{e^{K} \pm e^{-K}}{\sqrt{2}}.
 \end{equation}
For the present Hamiltonian (\ref{generalspin}), 
 the dual Hamiltonian is given by 
 \begin{equation}
 H^{*} = -J \sum_{C^{*}} \prod_{\xv \in \partial C^{*}} S_{\xv},
 \end{equation}
 which is of the same form as equation (\ref{generalspin}). $C^{*}$ is the
 dual element of $C$ and has dimension $d-r$.
 Next we derive the dual expression of the partition function
 for general dual pairs.
 The partition function as a function of $u$
 has the following property,
\begin{equation}
 \label{o_d}
  {\cal Z}_{\rm orig} \{ u_1(K), u_{-1}(K) \}
 = 2^a {\cal Z}_{\rm dual} \{ u^{*}_{1}(K) ,u^{*}_{-1}(K) \},
 \end{equation}
 Here ${\cal Z}_{\rm orig}$
 is the partition function of the
 original model and ${\cal Z}_{\rm dual}$  is for the dual model.
 The symbol $a$ is a constant
 determined by the numbers of the elements on the lattice.
 (See appendix.) 

 We give several examples of dual pairs below, which would be 
 helpful to understand the definition of the dual transformation.
 The arrows indicate duality relations.
 \begin{itemize}
 \item $(d=2,r=1)$ 2$d$ Ising model on the square lattice $\leftrightarrow$ 2$d$
       Ising model on the square lattice (self-dual)
 \item $(d=2,r=1)$ 2$d$ Ising model on the triangular lattice
       $\leftrightarrow$ 2$d$ Ising model on the hexagonal lattice
 \item $(d=4,r=2)$ 4$d$ $Z_2$ lattice gauge model on the hypercubic
       lattice $\leftrightarrow$ 4$d$ $Z_2$ lattice gauge model on the hypercubic lattice (self-dual)
 \item $(d=3,r=1)$ 3$d$ Ising model on the cubic lattice $\leftrightarrow$ 3$d$
       $Z_2$ lattice gauge model on the cubic lattice
 \end{itemize}
So far the discussions have not been restricted to self-dual models.

 For self-dual cases, ${\cal Z}_{\rm orig}$ and ${\cal Z}_{\rm dual}$
 are the same function and the prefactor $2^a$ on the right hand side 
 in (\ref{o_d}) becomes a trivial constant which is negligible in the
 thermodynamic limit (and is omitted in the following).
 Hence (\ref{o_d}) is simplified to 
 \begin{equation}
 {\cal Z} \{ u_1(K), u_{-1}(K) \} = {\cal Z}
  \{ u^{*}_{1}(K) ,u^{*}_{-1}(K) \},
 \end{equation}
where the symbol ${\cal Z}_{\rm orig/dual}$ is simplified to $\cal Z$.
 From this expression it is seen that $\cal Z$ is invariant under the exchange
 $u_{1} (K) \leftrightarrow u^{*}_{1} (K)$ and $u_{-1} (K) \leftrightarrow u^{*}_{-1} (K)$,
 which means
 self-duality of the partition function.
 The critical point of a self-dual model is obtained, 
 if it is unique, by the fixed-point condition for these Boltzmann factors,
 $u_{\pm 1}(K_c)=u^{*}_{\pm 1}(K_c)$, which yields 
 $K_c=\frac{1}{2} \ln (\sqrt{2}+1)$.
 It is clear that 
 this transition point is shared by the 2$d$ Ising model on the 
 square lattice, the 4$d$ $Z_2$ gauge model on
 the hypercubic lattice and their higher dimensional generalizations \cite{Wegner}.

\subsection{Duality of random models and conjecture
for the critical point}
 
 Let us introduce randomness. To investigate
 critical points of random systems with the aid of the duality formalism,
 we will utilize the technique of reference \cite{MNN} with some
 modifications.
 
 The random model treated here is a system
 with $Z_2$ variables and bimodal randomness.
 The Hamiltonian is written as
\begin{equation}
 \label{generalrandomspin}
 H = - J \sum_{C} \tau_{C} \prod_{\xv \in \partial C} S_{\xv}, 
 \end{equation} 
where $\tau_C$ is a quenched random variable dependent on
 each element $C$. $\tau_C$ takes the value $1$
 with probability $p$ and $-1$ with $1-p$. 
 To treat random systems, we employ the standard replica method.
 Let us consider the $n$-replicated system
 and define the averaged Boltzmann factor
 $x_{k}$ for an element $C$, which corresponds to
 the configuration
 $\prod_{\xv \in \partial C} S_{\xv} = 1$ in $n-k$ replicas
 and $-1$ in $k$ replicas.
 The explicit form of $x_{k}$ is
 \begin{equation}
 x_{k} (p,K) = p e^{(n-2k)K} + (1-p) e^{-(n-2k)K}.
 \end{equation}
The $n$-replicated partition function is, after average over randomness,
 a function of these Boltzmann factors,
 \begin{equation}
 \label{Zisotropy}
 [{\cal Z}^n]_{\rm{av}} \equiv  {\cal Z}_n \{ x_{0}(p,K), x_{1}(p,K),\ldots, x_{n}(p,K) \},
 \end{equation}
where $[\ ]_{\rm{av}}$ means random average. 

 We also define the dual Boltzmann
 factor $x_{k}^{*}(p,K)$ on the dual lattice.
 The explicit forms are obtained by the two-component
 Fourier transformation with the result
 \begin{eqnarray}
 \label{dualBoltzmann}
 x_{2k}^{*} (p,K) & = & 2^{-n/2} (e^{K} + e^{-K})^{n-2k} (e^{K} - e^{-K})^{2k}, \nonumber \\
 x_{2k+1}^{*} (p,K) & = & 2^{-n/2}(2p-1) (e^{K} + e^{-K})^{n-2k-1}
 (e^{K} - e^{-K})^{2k+1},
 \end{eqnarray}
where $k$ is a non-negative integer in the range $0\le 2k<2k+1\le n$.
 The partition function satisfies a generalization of (\ref{o_d}),
\begin{equation}
\label{5a}
 {\cal Z}_{n,{\rm orig}}\{ x_0, x_1,\cdots ,x_n\}
  =2^{\tilde{a}}{\cal Z}_{n,{\rm dual}}\{ x_0^*, x_1^*,\cdots ,x_n^*\},
\end{equation}
where $\tilde{a}$ is an appropriate constant.
Now we restrict our attention to the case where the system is self-dual
 when we remove randomness (e.g. 2$d$ $\pm J$ random bond Ising model on the
 square lattice). 
 Using (\ref{5a}) we can express the self duality
 of the $n$-replicated partition function for such a system as
 \begin{equation}
 \label{selfdual}
  {\cal Z}_n \{ x_{0}, x_{1}, \ldots, x_{n} \}
 =  {\cal Z}_n \{ x_{0}^{*},x_{1}^{*},\ldots, x_{n}^{*} \},
  \end{equation}
where an overall constant is neglected.
 Thus self duality is recognized by the fact
 that ${\cal Z}_n$ is invariant
 when the exchanges $x_k(p,K) \leftrightarrow x_k^{*}(p,K)$ for all $k$
 are performed
 simultaneously.

 It is in general impossible to identify the transition point from the fixed-point
 condition of the duality relation (\ref{selfdual}), unlike the non-random
 case because the fixed-point conditions of all the variables
 $x_0=x_0^*, x_1=x_1^*, \cdots ,x_n=x_n^*$ are not satisfied simultaneously.
 The authors of \cite{MNN,TN} nevertheless developed an argument leading
 to a conjecture that the fixed-point condition of the leading Boltzmann factor
\begin{equation}
 \label{selfdual2}
 x_{0}(p_c,K_c) = x_{0}^{*} (p_c,K_c)
 \end{equation}
can well be the most plausible candidate to give
 the exact transition point of the random system at least on the
 Nishimori line (NL) \cite{NL},
 $e^{-2K}=(1-p)/p$, where enhanced symmetry simplifies the system
 properties significantly.
 
 This prediction has been confirmed to be correct in the cases of $n=1, 2$
 and $\infty$ \cite{MNN}. It has also been shown that numerical results for
 the quenched limit $n\to 0$ \cite{numerical} agree very well with the
 conjectured value of $ p_c=0.889972...$ which is the solution to the
 formula obtained in the $n\to 0$ limit of $x_0=x_0^*$ on the NL:
 \begin{equation}
  -p \log_2 p-(1-p)\log_2 (1-p)=\frac{1}{2}.
 \end{equation}
 This conjecture also leads us to an interesting result that
 the multicritical points of the $2d$ $\pm J$
 random bond Ising model and the $4d$ random $Z_2$ lattice gauge model
 (random plaquette gauge model) are located at the same point
 on the $p$-$K$ plane \cite{TN}. This observation has also been confirmed numerically 
 by a recent study \cite{AIMK}.
 
 Determination of the multicritical point is also quite important
 from the standpoint of the quantum information theory, quantum memory in particular.
 To be specific,
 the value $1-p_c$ at the multicritical point
 of the $2d$ random bond Ising model
 is equivalent to the accuracy threshold of the 
 $2d$ toric code with perfect measurement \cite{Kitaev,DKLP},
 which is estimated to be $0.110028\ldots$
 from the above discussion.
 In addition, the value $1-p_c$ at the multicritical point of the $4d$
 random gauge model also gives the accuracy threshold of the 
 $4d$ toric code (or $3d$ code with imperfect measurement) \cite{TN},
 which is also determined as $0.110028\ldots$.


\section{$Z_2$ models}

With the knowledge of the previous section in mind, we proceed to
discussions on the non-self-dual cases with and without randomness.

\subsection{Duality for non-self-dual cases}

In this subsection, we develop an argument for duality of generic
non-self-dual $Z_2$ models.
It is clear that a simple duality relation like (\ref{o_d}) is not enough 
to determine the 
transition point. Nevertheless we show that, by generalizing the arguments
in the previous section,
one can still derive a relation between the transition points of 
a model and its dual.

First let us discuss non-random systems.
Consider the product of partition functions of the original and 
dual models with inverse temperatures $K_1$ and $K_2$, respectively. 
From (\ref{o_d}), one finds
 \begin{eqnarray}
 \hspace{-5mm}
 && {\cal Z}_{\rm orig} \{ u_1(K_1), u_{-1}(K_1) \} 
  {\cal Z}_{\rm dual} \{ u_{1}(K_2) ,u_{-1}(K_2) \} \nonumber \\
 \hspace{-5mm}
 &=& {\cal Z}_{\rm orig} \{ u^{*}_{1}(K_2), u^{*}_{-1}(K_2) \}
  {\cal Z}_{\rm dual} \{ u^{*}_{1}(K_1) ,u^{*}_{-1}(K_1) \},
 \end{eqnarray}
 which indicates that the product is invariant under the
  simultaneous exchange
  $u_{\pm 1}(K_1) \leftrightarrow u^{*}_{\pm 1}(K_2)$ and 
  $u_{\pm 1}(K_2) \leftrightarrow u^{*}_{\pm 1}(K_1)$.
 Hence, if there is a unique transition point $K_{1c}$
 (resp. $K_{2c}$) in the original (resp. dual) model, 
 the relation between two critical points   
 is expected to be given by 
 \begin{equation}
  u_{\pm 1}(K_{1c})u_{\pm 1}(K_{2c})
 = u^{*}_{\pm 1}(K_{1c}) u^{*}_{\pm 1}(K_{2c}),
 \end{equation}
which is invariant under the transformation above.
 One can verify that this is equivalent to 
 $e^{-2K_{2c}} = \tanh K_{1c}$, which gives the correct 
 relation between the two transition points.

 Now we move on to random systems. In this case, we can express
 the duality between two random models (\ref{5a})
 as follows,
\begin{eqnarray}
 \hspace{-1cm} && {\cal Z}_{n, \rm orig}
 \{ x_{0}(p_1,K_1), \ldots ,x_{n}(p_1,K_1) \}
  {\cal Z}_{n, \rm dual}\{ x_{0}(p_2,K_2), \ldots ,x_{n}(p_2,K_2) \}
 \nonumber \\
 \hspace{-1cm} &=& 
  {\cal Z}_{n, \rm orig}
          \{x^{*}_{0}(p_2,K_2),\ldots ,x^{*}_{n}(p_2,K_2) \}
  {\cal Z}_{n, \rm dual}\{ x^{*}_{0}(p_1,K_1), \ldots ,x^{*}_{n}(p_1,K_1) \},
 \end{eqnarray}
where $p_1,p_2$ and $K_1,K_2$ denote the probability of positive interaction
 and the inverse temperature for the
 original/dual models, respectively.
 Thus the product of the partition functions is invariant under
  the simultaneous exchange 
  $x_{k}(p_1,K_1) \leftrightarrow x^{*}_{k}(p_2,K_2)$ and 
  $x_{k}(p_2,K_2) \leftrightarrow x^{*}_{k}(p_1,K_1)$
 for all $k$.

 The argument developed so far naturally suggests that the relation between
 the critical points of the original and dual systems is given by the fixed-point
 condition of the leading Boltzmann factors at least on the NL.
 Explicitly, this condition reads
 \begin{equation}
 \label{selfdual3}
 x_{0}(p_{1c},K_{1c}) x_{0}({p_{2c}},{K_{2c}}) 
 = 
 x_{0}^{*}(p_{1c},K_{1c}) x_{0}^{*}(p_{2c},K_{2c}),
 \end{equation}
 in conjunction with the NL condition
 \begin{equation}
 \label{NL}
 e^{-2K_1} = \frac{1-p_1}{p_1},\hspace{5mm} e^{-2K_2} = \frac{1-p_2}{p_2}. 
 \end{equation}
 Equation (\ref{selfdual3}) with (\ref{NL})
 is written in terms of $p_{1c}$ and $p_{2c}$ as
\begin{equation}
 \label{dual_mc_n}
 \left( {p_{1c}}^{n+1} + (1-p_{1c})^{n+1} \right) 
   \left( {{p_{2c}}}^{n+1} + (1-{p_{2c}})^{n+1} \right)  = 2^{-n}.
 \end{equation}
 We expect that the two multicritical points
 are related by this equation.
 If the quenched ($n \rightarrow 0$) limit is taken, this yields the relation,
 \begin{equation}
 \label{dualcriticalpoint}
 H(p_{1c}) + H(p_{2c}) = 1,
 \end{equation}
where
\begin{equation}
 H(p) \equiv -p \log_2 p -(1-p) \log_2(1-p).
 \end{equation} 
 Therefore our main statement for the non-self-dual $Z_2$ model is the following:
 the two critical points on the NLs of the mutually-dual systems
 with quenched randomness are expected to be related by equation (\ref{dualcriticalpoint}).
  
 There are many reasons to believe that our conjecture expressed in
 (\ref{dual_mc_n}) and (\ref{dualcriticalpoint}) is exact. 
 Some of them are given in the rest of this paper.
 For simplicity the discussion in the following is restricted to the
 $2d$ random bond Ising models on the mutually-dual lattices such as
 the hexagonal and triangular lattices,
 though the conjecture applies quite generally to arbitrary systems
 described by the Hamiltonian (\ref{generalrandomspin}).

 Before closing the present subsection, we explain an explicit
 representation of the dual random models \cite{NN,MNN}, which is necessary in
 the following discussions.
 The ratios of dual Boltzmann factors (\ref{dualBoltzmann}) to $x_0^*$
 are
 \begin{eqnarray}
 x_{2k}^{*} / x_{0}^{*} & = & \tanh^{2k} K, \nonumber \\
 x_{2k+1}^{*} / x_{0}^{*} & = & (2p-1) \tanh^{2k+1} K.
 \end{eqnarray}
These ratios of Boltzmann factors are realized by a system with the following
 explicit Boltzmann factors written in terms of Ising spin variables
\begin{equation}
 \label{dualrep}
 A \exp \left( \tilde{K} (S^{(1)}+S^{(2)}+ \ldots + S^{(n)})
 + \tilde{K}_{p} S^{(1)}S^{(2)}\ldots S^{(n)} \right),
 \end{equation}
where $S^{(k)}$ is the product of Ising spin variables in the $k$th replica. 
 For example, in the case of the usual nearest neighbour interactions,
 $S^{(k)}$ stands for $S_i^{(k)}S_j^{(k)}$.
 $\tilde{K}$ and $\tilde{K}_{p}$ are defined by
\begin{equation}
 \label{dualKp}
 e^{-2 \tilde{K}} \equiv \tanh K, \ \ e^{-2 \tilde{K}_{p}} \equiv 2p-1 (\equiv \tanh K_{p}).
 \label{KKp}
 \end{equation}
 From the expression (\ref{dualrep}), the dual of the
 $\pm J$ random bond Ising model can be interpreted as the 
 model with the non-random Ising interaction in each replica
 and the interaction between replicas.
 If the condition of the NL, $\tilde{K}= \tilde{K}_{p}$, is imposed,
 this turns to
\begin{equation}
 \label{dualrep2}
 A \exp \left( \tilde{K}(S^{(1)}+S^{(2)}+ \ldots + S^{(n)} 
 + S^{(1)}S^{(2)}\ldots S^{(n)} ) \right).
 \end{equation}
%
 \subsection{Verification for $n=1$}

 Let us first show that the relation (\ref{selfdual3}) gives the exact answer even without
 the NL condition when $n=1$. 
 Equation (\ref{selfdual3}) is, for $n=1$,
 \begin{equation}
 \label{dualn1}
 \hspace{-2cm}
 ( p_c e^{K_{1c}} + (1-p_{1c}) e^{-K_{1c}} )
 ( {p_{2c}} e^{K_{2c}} + (1-{p_{2c}})
 e^{-K_{2c}} ) = \frac{1}{2} (e^{K_{1c}} + e^{-K_{1c}})( e^{K_{2c}}
 + e^{-K_{2c}}). 
 \end{equation}
>From the expression (\ref{dualrep}), it is found that
 the ${\pm J}$ random bond Ising model for $n=1$, averaged over randomness, is regarded
 as a non-random Ising model with coupling $\tilde{K} + \tilde{K}_{p}$
 on the dual lattice.
 For example,
 the random bond Ising model on the 2$d$ triangular lattice with parameters $p_1$ and 
 $K_1$ is equivalent to the non-random Ising model on the hexagonal lattice,
 whose inverse temperature
 $\hat{K}_1$ is given by
\begin{equation}
 \hat{K}_1 = \tilde{K}_1 + \tilde{K}_{p1},
 \end{equation} 
where $\tilde{K}_{p1}$ is defined by the second expression of (\ref{KKp}) with
 $p$ replaced by $p1$.
 Conversely, the random bond Ising model on the 2$d$ hexagonal
 lattice with parameters $p_2$ and $K_2$ corresponds to the non-random Ising model
 on the triangular lattice, whose inverse temperature
 $\hat{K}_2$ is
\begin{equation}
 \hat{K}_2 = \tilde{K}_2 + \tilde{K}_{p2}. 
 \end{equation} 
 As is well-known, the non-random Ising models on the 2$d$ triangular and the
 hexagonal lattices are mutually-dual, and two critical inverse temperatures 
 satisfy \cite{ItzyksonDrouffe}
\begin{equation}
 \label{dualpuren1}
 e^{-2 \hat{K}_{1c}} = \tanh \hat{K}_{2c}. 
 \end{equation}
 Using equation (\ref{dualKp}) we can confirm the equivalence
 between equations (\ref{dualn1}) and (\ref{dualpuren1}).
 Note that the condition of the NL is not used,
 and this equivalence holds everywhere on the phase boundary.

 \subsection{Verification for $n=2$}

 Next we discuss the two-replica case.
 Equation (\ref{selfdual3}) for $n=2$ is
\begin{equation}
 \label{dualn2}
 \hspace{-2.7cm}
 ( p_{1c} e^{2K_{1c}} + (1-p_{1c}) e^{-2K_{1c}} )( {p_{2c}} e^{2K_{2c}}
 + (1-{p_{2c}}) e^{-2K_{2c}} )
 = \frac{1}{4} (e^{K_{1c}} + e^{-K_{1c}})^{2}( e^{K_{2c}} + e^{-K_{2c}})^{2}. 
 \end{equation}
 In the rest of this subsection, 
 we restrict our attention onto the NL. 

 Due to the dual representation 
 (\ref{dualrep2}) for $n=2$, we know that the random bond Ising model
 is equivalent to the non-random four-state Potts model on the dual lattice;
 four states are constructed by
 the direct  product of Ising factors in two replicas  $S^{(1)}=\pm 1,
 S^{(2)}= \pm 1$, and the dual Boltzmann factor is $A e^{3 \tilde{K}}$ for
 $S^{(1)}=S^{(2)}=1$ and $A e^{ -\tilde{K}}$ otherwise. Therefore
 the system corresponds to the four-state Potts model on the dual lattice
 with coupling $2 \tilde{K}$, namely,
 \begin{equation}
 \beta H = -2 \tilde{K} \left( 2 \delta_{S^{(1)},1} \delta_{S^{(2)},1} -1 \right)-\tilde{K}.
 \end{equation}
 For example, the random bond Ising
 model on the triangular lattice is equivalent
 to the non-random four-state Potts model on the hexagonal lattice and vice versa.
 Here we denote the parameters of
 the random bond Ising model on the triangular lattice by $p_1,K_1$ 
 and that on the hexagonal lattice by $p_2, K_2$.
 Then the inverse temperatures of the corresponding four-state Potts models
 $\hat{K}_1, \hat{K}_2$ are given by
 \begin{equation}
 \hat{K}_1 = 2 \tilde{K}_1, \hspace{1cm} \hat{K}_2 = 2 \tilde{K}_2,
 \end{equation}
 as mentioned above. The dual coupling $\tilde{K}_1$ (or $\tilde{K}_2$)
 is defined by equation (\ref{dualKp}).

 The inverse critical temperatures of the four-state Potts models
 on mutually-dual lattices satisfy \cite{ItzyksonDrouffe}
 \begin{equation}
 \label{dualpuren2}
 e^{-2 \hat{K}_{2c}} = \frac{ e^{\hat{K}_{1c}} - e^{-\hat{K}_{1c}}  }
 {e^{\hat{K}_{1c}} + 3  e^{-\hat{K}_{1c}}}.
 \end{equation}
 It is straightforward to show that equation (\ref{dualn2}) is equivalent to
 equation (\ref{dualpuren2}) under the NL condition (\ref{NL}).

 \subsection{The limit $n \rightarrow  \infty$}

 Next is an argument for the $n \rightarrow \infty$ limit.
 The partition function for arbitrary $n$ is expressed as
 \begin{equation}
 {\cal Z}_{n} = [ {\cal Z}^{n}]_{\rm av} = [ e^{-n \beta N f(K)}]_{\rm av},
 \end{equation}
 where $N$ is the number of sites
 and $f(K)$ is the free energy for the inverse temperature $K$.
 If the replica number $n$ is taken to be quite large,
 the averaged partition function is
 dominated by the configuration of the smallest value of $f(K)$,
 which is realized by the perfect ferromagnetic bond configuration and its
 gauge equivalents. Using this argument ${\cal Z}_{n}$ can be approximated as
 \begin{equation}
 {\cal Z}_{n} \approx e^{-n \beta N f_0(K)},
 \end{equation} 
 where $f_{0}(K)$ is the free energy of the non-random ferromagnetic system.
 Therefore the random bond Ising system for $n \rightarrow
 \infty$ is interpreted as the non-random ferromagnetic Ising system
 (and its critical point is denoted by $K_{1c}$).
  
 We also consider the random bond Ising model on the dual lattice, and
 for $n \rightarrow \infty$ 
 we can regard it as the non-random Ising model 
 with the critical point ${K_{2c}}$.
 These two critical points are related by the well-known equation,
 \begin{equation}
 \label{dualninfty}
 e^{-2 K_{2c}} = \tanh K_{1c}.
 \end{equation}
 If we consider the $n \rightarrow \infty$ limit of
 equation (\ref{selfdual3}) combined with the NL condition (\ref{NL}),
 we obtain
 \begin{equation}
 e^{K_{1c} + K_{2c}} = \frac{(e^{K_{1c}} + e^{-K_{1c}}) (e^{K_{2c}}
 + e^{-K_{2c}})}{2}, 
 \end{equation}
 which is equivalent to equation (\ref{dualninfty}). 
 
 \subsection{Numerical evidence}

 We have checked the relation (\ref{dualcriticalpoint}) in the quenched limit numerically.
 We executed Monte Carlo simulations for the $2d$ random bond Ising 
 models on the hexagonal and the triangular lattices,
 a dual pair.
 To observe the criticality, we make use of the 
 non-equilibrium relaxation (NER) method \cite{NER}, which yields the
 power-law behaviour of decreasing magnetization
 with Monte Carlo steps on criticality.
 For this method we prepare all-up spins as the initial state
 and let them relax in each Monte Carlo step.
 In order to identify the multicritical point,
 we choose the parameters of the system to be on the NL
 and vary the parameter $p$ (and $K$ is also varied accordingly). 

 The results are shown in figure 1.
 From these results, the locations of the multicritical points
 are estimated at $p_{1c} = 0.930(5)$
 (or $0.347 <H(p_{1c})< 0.384$)
 for the hexagonal lattice and
 at $p_{2c} = 0.835(5)$ (or $0.634< H(p_{2c}) < 0.658$)
 for the triangular lattice.
 The two $H$'s sum up to
 $0.981<H(p_{1c})+H(p_{2c})<1.042$, a consistent result
 with our conjecture in equation (\ref{dualcriticalpoint}).
\begin{figure}
\begin{center}
\includegraphics[width=15.7cm]{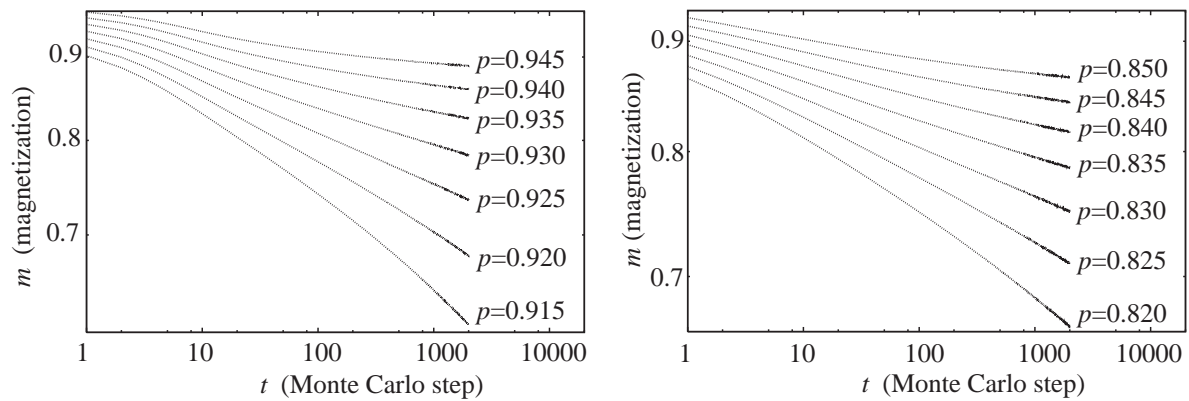}
\label{NERtrihex}
\caption{\small Relaxation of magnetization near the multicritical points of
 the random bond Ising model
 on the hexagonal (left) and the triangular (right) lattices from NER analysis.
 We prepared $L\times L (L=10^3)$ spins and averaged the results over $200$
 samples. The system at criticality is expected to yield a straight
 line in this log-log plot. }
\end{center}
\end{figure}

\subsection{Random models in $3d$}

 As already noted in section 2, the random bond Ising model 
 and the random $Z_2$ lattice gauge model on the $3d$ cubic lattice
 are mutually dual. 
 Thus we expect that the multicritical points of these two models
 will satisfy equation (\ref{dualcriticalpoint}). 
 This conjecture can be confirmed by numerical simulations of these $3d$ systems.

 In fact the multicritical points of these two models have already been
 estimated.
 For the $3d$ random bond Ising model the multicritical point is
 estimated to be $p_{1c} = 0.7673(3)$ \cite{IOK}
 which yields $H(p_{1c})\approx 0.783$.
 For the $3d$ random gauge model
 it is $p_{2c} \approx 0.967$  \cite{OAIM},
 giving $H(p_{2c}) \approx 0.209$.
 From these results the sum $H(p_{1c})+H(p_{2c})$ is
 about $0.992$,  a reasonable value in view of our expectation (\ref{dualcriticalpoint}).

 From the viewpoint of the quantum information theory, the accuracy
 threshold of the $2d$ toric code
 with imperfect measurement can be
 determined by the value $1-p_c$ at the multicritical point of the $3d$ 
 random gauge model \cite{Kitaev,DKLP}.
 Therefore it is an advantage of the present analysis that the duality gives
 an alternative way to determine the accuracy threshold
 when we utilize the result of the $3d$ random bond Ising model.  
 From the result
 of the direct numerical analysis of the $3d$ random gauge model
 \cite{OAIM} and the duality argument combined
 with the numerical result of the $3d$ random bond Ising model \cite{IOK},
 the accuracy threshold is estimated to be about $0.03 \sim 0.035$.

 \subsection{2$d$ anisotropic random bond Ising model and self duality}

 In this and the next subsections we discuss the $2d$ random bond 
 Ising model with anisotropic disorder.
 This model is {\it not} a non-self-dual model, but the structure 
 of the problem is similar to the mutually-dual case.
 We show that the duality formalism can be applied to this system and
 gives a conjecture for the critical points.
 It should also be noted that the critical point of this model 
 gives the accuracy threshold of the $1d$ quantum repetition code
 with imperfect measurement \cite{DKLP}.

 First we review the duality of the anisotropic non-random Ising model on
 the $2d$ square lattice. The Hamiltonian is
 \begin{equation}
 \label{nonrandomising}
 H= -J_h \sum_{\langle ij \rangle \in C_h} S_i S_j 
 - J_v \sum_{\langle ij \rangle \in C_v} S_i S_j ,
 \end{equation}
 where $J_h$ and $J_v$ are uniform coupling constants. The symbol $C_h$ denotes
 the set of horizontal bonds and $C_v$ is for vertical ones. 
 The partition function is
 \begin{equation}
 {\cal Z} = \sum_{\{S_{i}= \pm 1\}} \prod_{\langle ij \rangle \in C_h}
 u_{\scriptscriptstyle S_{ij}}(K_h) 
 \prod_{\langle ij \rangle \in C_v} u_{\scriptscriptstyle S_{ij}}(K_v) ,
  \label{bondsumising}
 \end{equation}
 where $K_{h,v} \equiv \beta J_{h,v}=J_{h,v}/kT$ and $S_{ij} \equiv S_i S_j$.
 The symbol $u$ is the Boltzmann factor for the bond between $i$ and $j$,
 \begin{equation}
 u_{\scriptscriptstyle S_{ij}}(K_{h,v}) \equiv \exp ( K_{h,v} S_{ij} ).
 \end{equation}

 Next we define the dual Boltzmann factor by the binary Fourier transformation,
 \begin{equation}
 u^{*}_{\pm 1}(K_{h,v}) \equiv \frac{u_1(K_{h,v}) \pm u_{-1}(K_{h,v})}{\sqrt{2}} 
 = \frac{e^{K_{h,v}} \pm e^{-K_{h,v}}}{\sqrt{2}}.
 \end{equation} 
 Using these Boltzmann factors, we can express the duality of
 partition function as
 \begin{equation}
 \label{dualZising}
 \hspace{-2cm}
 {\cal Z} \{ u_{1}(K_{h}),u_{-1}(K_{h}),u_{1}(K_{v}),u_{-1}(K_{v}) \} =
 {\cal Z} \{ u_{1}^{*}(K_{v}),u_{-1}^{*}(K_{v}),
  u_{1}^{*}(K_{h}),u_{-1}^{*}(K_{h}) \}.
 \end{equation}
 In this expression the Boltzmann factors of
 the vertical and horizontal bonds are exchanged
 because a vertical bond is mapped to a horizontal bond
 on the dual lattice and vice versa.

 It is obvious that the partition function is invariant 
 under the exchange
  $u_{\pm 1}(K_h) \leftrightarrow u^{*}_{\pm 1}(K_v)$ and 
  $u_{\pm 1}(K_v) \leftrightarrow u^{*}_{\pm 1}(K_h)$, which
 is similar to the case of mutually-dual non-random systems.
 The critical points are determined by the equation,
 \begin{equation}
 \label{dualnonrandomani}
  u_{\pm 1} (K_{h}) u_{\pm 1} (K_{v}) =
  u_{\pm 1}^{*} (K_{v}) u_{\pm 1}^{*} (K_{h}),
 \end{equation}
 which yields $e^{-2K_v} = \tanh K_h$. 

 Next we study the system with randomness. The Hamiltonian is
 \begin{equation}
 \label{randomising}
  H= -J_v \sum_{\langle ij \rangle \in C_v} \tau^{v}_{ij} S_i S_j - J_h \sum_{\langle ij \rangle \in C_h} \tau^{h}_{ij} S_i S_j ,
 \end{equation}
 where $\tau^{v,h}$ are random variables which depend on each
 bond and
 obey the probability distribution,
 \begin{equation}
 P(\tau^{h,v}_{ij}) = p_{h,v} \delta(\tau^{h,v}_{ij} - 1) + (1-p_{h,v}) \delta(\tau^{h,v}_{ij} + 1),
 \end{equation} 
 for horizontal$(h)$ or vertical$(v)$ bond.
 The averaged partition function is a function of averaged
 Boltzmann factors $x_{k} (p_h,K_h)$ and $x_{k}(p_v,K_v)$,
 \begin{equation}
 \label{Zanisotropy} 
 \hspace{-1cm}
  [{\cal Z}^n]_{\rm{av}} \equiv 
 {\cal Z}_n \{ x_{0}(p_h,K_h),\ldots, x_{n}(p_h,K_h), x_{0}(p_v,K_v),\ldots, x_{n}(p_v,K_v) \}.
 \end{equation}
 We also define the dual averaged Boltzmann factors
 in the same way as in section 2,
 \begin{eqnarray}
 \hspace{-2cm} x_{2k}^{*}(p_{h,v},K_{h,v}) & = & 2^{-n/2} (e^{K_{h,v}} + e^{-K_{h,v}})^{n-2k} (e^{K_{h,v}} - e^{-K_{h,v}})^{2k}, \nonumber \\
 \hspace{-2cm} x_{2k+1}^{*}(p_{h,v},K_{h,v}) & = & 2^{-n/2} (2p_{h,v}-1) (e^{K_{h,v}} + e^{-K_{h,v}})^{n-2k-1} (e^{K_{h,v}} - e^{-K_{h,v}})^{2k+1}.
 \end{eqnarray}
 Using $x_{k}$ and $x_{k}^*$, we can express the duality of the $n$-replicated
 partition function,
 \begin{eqnarray}
 \hspace{-1cm} && {\cal Z}_n \{x_{0}(p_h,K_h), \ldots, x_{n}(p_h,K_h), x_{0}(p_v,K_v) , \ldots, x_{n}(p_v,K_v) \} \nonumber \\
 \hspace{-1cm} &=&  {\cal Z}_n \{x_{0}^{*}(p_v,K_v), \ldots, x_{n}^{*}(p_v,K_v), x_{0}^{*}(p_h,K_h) , \ldots, x_{n}^{*}(p_h,K_h) \},
  \end{eqnarray}
 which is invariant under the simultaneous exchange
 $x_{k}(p_h,K_h) \leftrightarrow x^{*}_{k}(p_v,K_v)$ and
 $x_{k}(p_v,K_v) \leftrightarrow x^{*}_{k}(p_h,K_h)$ for all $k$.
 The Boltzmann factors of the vertical and horizontal bonds
 should be exchanged as in the non-random case.

 From the argument above, we make
 a conjecture for the critical points on the NLs, 
 \begin{eqnarray}
 \label{randomanicritical}
 x_{0}(p_h,K_h) x_{0}(p_v,K_v) = x_{0}^{*}(p_h,K_h) x_{0}^{*}(p_v,K_v),
 \end{eqnarray}
 from the analogy with the non-random case (\ref{dualnonrandomani})
 or the non-self-dual random case (\ref{selfdual3}).
In this system, the NLs are defined for the horizontal and vertical bonds,
respectively, by
\begin{equation}
 e^{-2K^{h}} = \frac{1-p_h}{p_h}, \hspace{5mm} e^{-2K^{v}} = \frac{1-p_v}{p_v}.
 \label{Khv}
\end{equation} 
 We should note that the original system
 has four parameters $p_h,K_h,p_v,K_v$ and the two conditions of the NLs
 reduce the number of independent variables to two.
 Consequently, the conjecture (\ref{randomanicritical})
 combined with the conditions
 of the NLs (\ref{Khv}) does not fix the multicritical point on the phase diagram
 but determines the location of the ``critical line''. This fact is
 favourable for numerical verification of our conjecture
 because one parameter can be chosen freely even on criticality and on
 the NLs.

\subsection{Verification for anisotropic system}

 We verify the validity of the conjecture in
 equation (\ref{randomanicritical}). For this purpose the following property 
 is useful; the form of equation (\ref{randomanicritical}) is
 completely the same as equation (\ref{selfdual3})
 when the parameters are replaced as
 \begin{equation}
 \label{replacement}
 \{ p_1,K_1,p_2,K_2 \} \rightarrow \{ p_h,K_h,p_v,K_v \}. 
 \end{equation}
 Using this correspondence
 we can check the conjecture (\ref{randomanicritical})
 by the same argument as in mutually-dual systems
 with $n=1,2$ and $n \rightarrow \infty$.

\begin{itemize}
\item{$n=1$}:
 The anisotropic random bond Ising model is
 equivalent to the anisotropic non-random Ising model,
 and equation (\ref{dualpuren1}) turns to the critical condition for
 the anisotropic non-random model when
 the replacement (\ref{replacement}) is used.

\item{$n=2$}:
The system is equivalent to the anisotropic non-random four-state Potts model.
 Using the replacement (\ref{replacement}),
 equation (\ref{dualpuren2}) becomes the critical condition for
 the anisotropic Potts model.

\item{$n \rightarrow \infty$}:
The system is equivalent to the anisotropic non-random Ising model and
the critical point satisfies equation (\ref{dualninfty}) using the correspondence
 (\ref{replacement}).

\end{itemize}

 If we consider the quenched limit $n \rightarrow 0$,
 we obtain the relation,
 \begin{equation}
 \label{anicriticalpoint}
 H(p_h) + H(p_v) = 1,
 \end{equation} 
 by the same argument as in mutually-dual systems (\ref{dualcriticalpoint}).
 Here we take $p_h,p_v$ as the two independent variables
 and eliminate $K_{h,v}$ using the conditions of the NLs.
 We expect that $p_h$ and $p_v$ will satisfy equation (\ref{anicriticalpoint})
 on criticality.

 We have checked equation (\ref{anicriticalpoint}) numerically.
 The result is shown in figure 2.
\begin{figure}
\begin{center}
  \includegraphics[width=7cm]{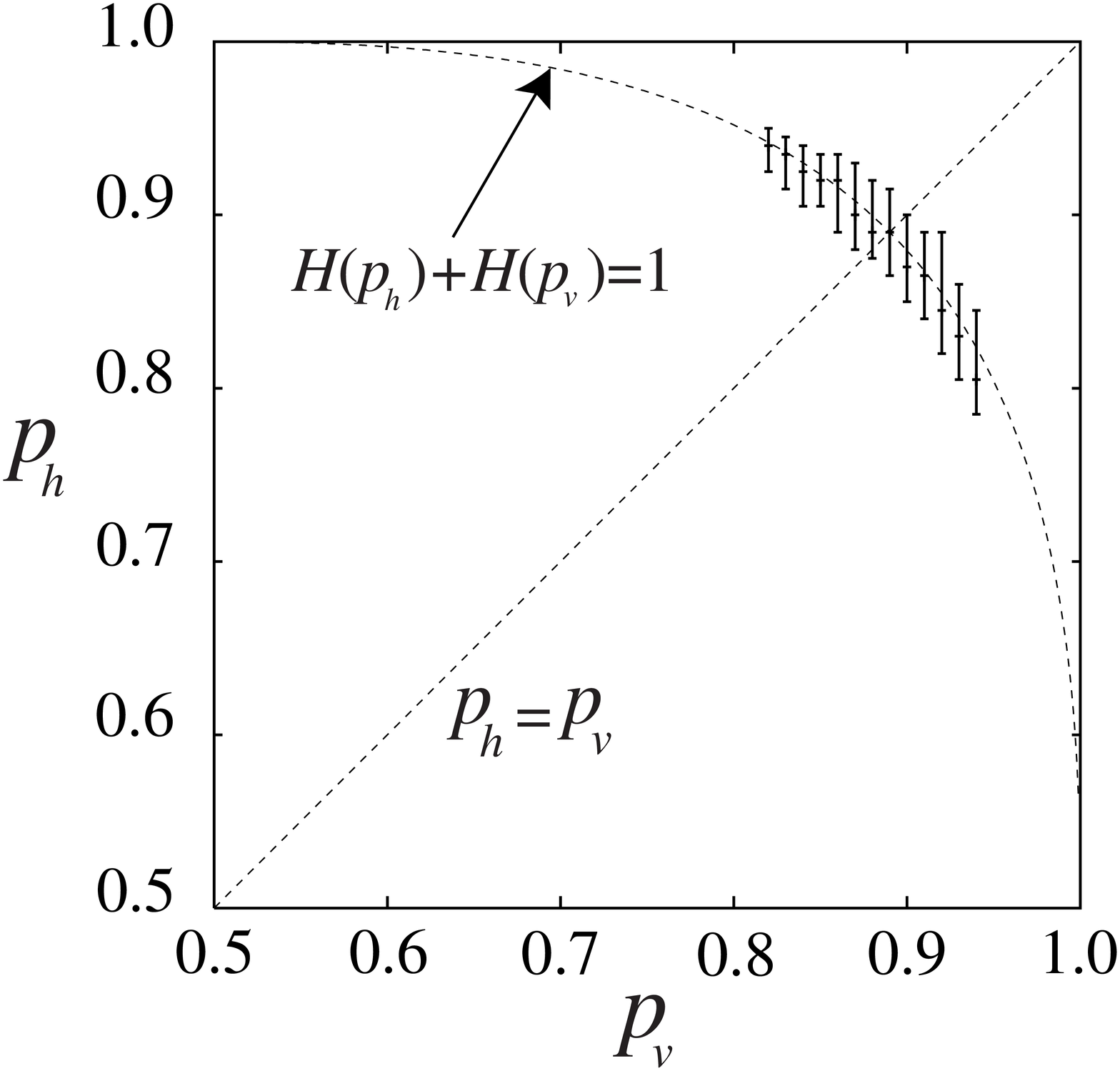}
\label{NERani}
\caption{\small Results by the NER for the anisotropic random bond Ising model.
 We simulated systems with up to $500 \times 500$ spins under
 all-up initial configuration. The results are averaged over $200$
 samples. For executing numerical calculation, we fixed $p_v$ and scanned $p_h$
 in order to search the critical point.
 The critical points obtained are shown with error bars.
 The curve $H(p_h)+H(p_v)=1$ lies within the error bars for all $p_v$. }
\end{center}
\end{figure}
 The locations of the critical points
 expected from equation (\ref{anicriticalpoint})
 are in reasonable agreement with numerical results.
 Using the NER with ordered and disordered initial
 configurations, we also checked that the result 
 for $p_v$ very close to $1$ (or $p_h \sim 1$)
 does not show discrepancy from the conjecture,
 which is not shown in the figure because the critical point
 could not be determined with sufficient precision
 due to rigid spin domains along one axis. 
 Thus we conclude that our conjecture is consistent
 with the numerical result for the anisotropic system as well.
 
 We are allowed to interpret (\ref{anicriticalpoint}) as
 the accuracy threshold of the $1d$ quantum repetition code,
 given by the value of $1-p_h$ under the imperfect
 measurement rate $1-p_v$. 


\section{$2d$ $Z_q$ model}

In this section we consider a generalization of the formalism to a
multi-valued spin model, the random $Z_q$ model.
The $Z_q$ model has $q$ states per site.
In the non-random case, this system in two dimensions has both aspects of the
ferromagnetic Ising model (i.e. with ferromagnetic order) and the XY model
(i.e. with the Kosterlitz-Thouless (KT) order) for $q$ not too small
\cite{JoseKadanoffKirkpatrickNelson1977,EPS1979}.
The authors of reference \cite{EPS1979} showed that this model has three
phases for $q\ge 5$, disordered, KT-like, and ferromagnetic, if there
exists a KT phase in the continuum limit $q\to\infty$.
Duality has been used to relate the two transition points.

The objective of the present section is to generalize the duality arguments
in the previous section and apply them to the randomized version of 
the $Z_q$ model. As far as the authors know, 
this type of models have not been studied analytically in detail in the literature. 

\subsection{Duality and analysis of the transition points}
\label{duality}

We consider the random chiral $Z_q$ model, 
for which the partition function is written in the form,
\begin{equation}
 {\cal Z} = \sum_{\{k\}}\prod_{\langle ij \rangle } 
      e^{V_K(k_{i}-k_{j}-l_{ij})},
\end{equation}
where $V_K(k)$ is the interaction which satisfies cyclic
condition, $V_K(k+q)=V_K(k)$.
Here $i,j$ are the site indexes, $k$ is the state variable which   
takes $q$ values, $0,1,\cdots,q-1$, with $q\in \mathbb{Z}_+ = \{1,2,\cdots\}$
and $l$ is the random variable on each bond which also takes 
$q$ values, $0,1,\cdots,q-1$. 
The probability that a random variable $l$ on a bond takes the value 
$l$ is denoted by $p_l$ ($\sum_l p_l =1$).
A collection of them is denoted by $\{p\}(=\{p_0,p_1\ldots p_{q-1}\})$
 in the sequel.
The summation $\sum_{\{k\}}$ is taken over state variables on all
sites and the product $\prod_{\langle ij \rangle}$ is 
taken over all bonds.
Only nearest neighbour interactions are assumed to exist on the
square lattice in the present section.

Our discussions are mostly for the $Z_q$ version of the 
Villain model \cite{Villain1975},
for which the Boltzmann factor is given by
\begin{equation}
 e^{V_K(k)}
 = \sum_{m=-\infty}^{\infty} e^{-\frac{K}{2}(\frac{2\pi}{q}k-2\pi m)^2}
  \left(=\frac{1}{\sqrt{2\pi K }}\sum_{l\in \mathbb{Z}}
         e^{-l^2/(2K)+2\pi i lk/q}\right).
\end{equation}

We generalize the duality arguments to the random $Z_q$ model.
Let us average the $n$-replicated partition function.
The resulting function ${\cal Z}_n$
can be written in terms of the $q^n$ variables (local Boltzmann factors for neighbouring
spin pairs),
\begin{equation}
 \chi_{k_1,\cdots,k_n}(\{p\},K)
 \equiv
 \sum_l p_l e^{V_K(l+k_1)+\cdots +V_K(l+k_n)},
\end{equation}
where $k_i=0,1,\cdots,q-1$ ($i=1,2,\cdots,n$).
The variable $\chi_{k_1,\cdots,k_n}(\{p\},K)$ is the generalization of the
Boltzmann factor $x_k(p,K)$ which appeared in section 2.
If we set $q=2$ and $V_K(0)=K,V_K(1)=-K$, we obtain
the relation between the present and previous Boltzmann factors,
\begin{equation}
 \label{chi}
 \hspace{-5mm}
 \chi_{k_1,\cdots,k_n} (\{p \},K) = x_{k}(p,K) ,
 \hspace{5mm} {\rm where} \hspace{3mm} k=\sum_{m=1}^n k_m  \hspace{5mm} (k_m=0,1).
\end{equation}   

We can define the dual model, for which the partition 
function can be written in terms of the dual variables.
They are defined by $q$-value Fourier transformations \cite{WuWang}
\begin{equation}
\label{chi_st}
\hspace{-1cm}
 \chi_{k_1,\cdots,k_n}^* (\{p\},K)
 \equiv q^{-n/2}
 \sum_l p_l \sum_{m_1} \omega^{k_1 m_1} e^{V_K(m_1+l)} \cdots
 \sum_{m_n} \omega^{k_n m_n} e^{V_K(m_n+l)},
\end{equation}
where $\omega=e^{2\pi i/q}$. 
A remark is in order. 
The variables $\chi_{k_1,\cdots,k_n}$ are already defined at the end of 
section 2 of reference \cite{MNN} in a different form from 
the ones given above. The expressions (74)-(79) in 
\cite{MNN} are not in general valid and should be 
replaced by (\ref{chi}) and (\ref{chi_st}).

Now we define
\begin{eqnarray}
 \chi_{\rm O}(\{p\},K) 
 & \equiv & \chi_{0,\cdots,0}(\{p\},K)  =
 \sum_{l=0}^{q-1} p_l e^{n V_K(l)},
 \nonumber \\
 \chi_{\rm O}^*(\{p\},K) & \equiv & \chi_{0,\cdots,0}^*(\{p\},K)
 = q^{-n/2} \sum_{l=0}^{q-1} p_l \left(\sum_{k=0}^{q-1} e^{V_K(k+l)}\right)^n,
\end{eqnarray}
and apply the same arguments as in the previous section to identification
of transition points of the $Z_q$ model.
If uniqueness of the critical point is assumed,
generalization of (\ref{selfdual2}) gives
\begin{equation}
\label{SDp}
 \chi_{\rm O}(\{p\},K) = \chi_{\rm O}^*(\{ p\},K).
\end{equation}
Remember that the non-random model does not have a unique transition point 
for $q\ge 5$ \cite{EPS1979}. 
Rather, two transition points are related to each other by the 
duality relation.
Therefore let us consider the case where there exist two transition 
points for the random case.
If there are two critical points $K_1,K_2$ for a given $\{p\}$, 
they may satisfy
\begin{equation}
\label{2CPp}
 \chi_{\rm O}(\{p\},K_1) \chi_{\rm O}(\{p\},K_2) 
 = \chi_{\rm O}^*(\{p\},K_1) \chi_{\rm O}^*(\{p\},K_2).
\end{equation}

Let us take the quenched ($n\to 0$) limit of these relations.
When uniqueness is assumed ($K_1=K_2$), the condition reads
\begin{equation}
\label{SDq}
 -\sum_{l=0}^{q-1} p_l 
 \log_q \left( \frac{e^{V_K(l)}}{\sum_{k=0}^{q-1}e^{V_K(k+l)}} \right)
 = 
 \frac{1}{2}.
\end{equation}
If there are two critical points, this is replaced by 
\begin{equation}
\label{2CPq}
-\sum_{l=0}^{q-1} p_l 
\log_q \left( \frac{e^{V_{K_1}(l)}}{\sum_{k=0}^{q-1}e^{V_{K_1}(k+l)}} \cdot 
            \frac{e^{V_{K_2}(l)}}{\sum_{k=0}^{q-1}e^{V_{K_2}(k+l)}} \right)
= 1.
\end{equation}

If we consider the non-random case, where $p_0=1,p_l=0(l\neq 0)$,
$\chi_{\rm O}$ and $\chi_{\rm O}^*$ read
\begin{eqnarray}
 \hspace{-1cm}
 \chi_{\rm O} (\{1,0\ldots\},K)  &=& \sum_{m\in \mathbb{Z}}
 e^{-\frac{K}{2}(2\pi m)^2}, \\
 \hspace{-1cm}
 \chi_{\rm O}^*(\{1,0\ldots\},K) &=& q^{-1/2}\sum_{k=0}^{q-1} \sum_{m\in \mathbb{Z}} 
      e^{-\frac{K}{2}(\frac{2\pi}{q}k-2\pi m)^2}
    = \frac{1}{\sqrt{2\pi K q}}\sum_{m\in \mathbb{Z}} e^{-\frac{q^2 m^2}{2K}}.
\end{eqnarray}
If there are two critical points,  equation (\ref{2CPp}) gives the relation between them,
\begin{equation}
 K_1 K_2 = \frac{q^2 }{4\pi^2}.
 \label{K1K2Zq}
\end{equation}
This agrees with the correct relation \cite{EPS1979}.
However, it is in general not expected that the condition (\ref{2CPp}) 
determines the whole shape of the phase boundary in the phase diagram
of the random $Z_q$ model.
Restriction to the NL, where enhanced symmetry helps us to predict various
exact results, is more likely to lead to reliable results.

\subsection{Duality on the NL}

Consider a specific choice of $p_l$,
\begin{equation}
 p_l(K_p) = \frac{e^{V_{K_p}(l)}}{\sum_{k=0}^{q-1}e^{V_{K_p}(k)}},
\end{equation}
which enables us to obtain exact results using gauge symmetry under
the NL condition \cite{ON93},
\begin{equation}
 K = K_p. 
\end{equation}
The above set $\{p_l\}$ on the NL is denoted by 
$\{p_{\scriptscriptstyle\rm NL}\}$.
As in the previous section, we try to identify the relation between two
transition points on the NL using the condition (\ref{2CPp}).

Let therefore $\chi_{\rm O}^{\ \scriptscriptstyle\rm NL}(K),
 \chi_{\rm O}^{* {\scriptscriptstyle\rm ~NL}}(K)$
 denote the variables on the NL,
\begin{eqnarray}
 \chi_{\rm O}^{\ \scriptscriptstyle\rm NL}(K) 
 &\equiv & \chi_{\rm O}(\{p_{\scriptscriptstyle\rm NL}\},K)
 =
 \sum_{l=0}^{q-1} p_l(K) e^{n V_K(l)},
 \nonumber \\
 \chi_{\rm O}^{* {\scriptscriptstyle\rm ~NL}}(K) 
 & \equiv & \chi_{\rm O}^*(\{p_{\scriptscriptstyle\rm NL}\},K) =
 q^{-n/2} \sum_{l=0}^{q-1} p_l(K) \left(\sum_{k=0}^{q-1} e^{V_K(k+l)}\right)^n.
\end{eqnarray}
Our conjecture for the replicated systems is the following.
If uniqueness of the critical point is assumed,
the location of it is determined by 
\begin{equation}
\label{SDnl}
 \chi_{\rm O}^{\ \scriptscriptstyle\rm NL}(K) = \chi_{\rm O}^{* {\scriptscriptstyle\rm ~NL}}(K).
\end{equation}
If there are two critical points $K_1,K_2$ on the NL, they satisfy
\begin{equation}
\label{2CPnl}
 \chi_{\rm O}^{\ \scriptscriptstyle\rm NL}(K_1)
 \chi_{\rm O}^{\ \scriptscriptstyle\rm NL}(K_2)
 =  \chi_{\rm O}^{* {\scriptscriptstyle\rm ~NL}}(K_1)
   \chi_{\rm O}^{* {\scriptscriptstyle\rm ~NL}}(K_2).
\end{equation}

We expect these relations to hold
even in the quenched ($n\to 0$) limit.
In this limit, the condition reads, corresponding to equation (\ref{SDnl}),
\begin{equation}
\label{SDnlq}
 -\sum_{l=0}^{q-1} p_l(K)\log_q p_l(K) = \frac{1}{2},
\end{equation}
and to equation (\ref{2CPnl})
\begin{equation}
\label{2CPnlq}
 -\sum_{l=0}^{q-1} \left( p_l(K_1)\log_q p_l(K_1)+p_l(K_2)\log_q p_l(K_2)\right) = 1.
\end{equation}
It is interesting to note that these conditions are written in 
terms of the entropy function. 
This fact is suggestive of some underlying structure behind
the scene, but we do not have a clear interpretation of this 
fact at the moment.

It is convenient to define
\begin{equation}
 F_K(x) \equiv \sum_{m=-\infty}^{\infty} e^{-\frac{K}{2}(x-2\pi m)^2}
\end{equation}
to facilitate more compact expressions for duality relation.
Clearly, we have
\begin{equation}
 e^{V_K(k)} = F_K\left(\frac{2\pi k}{q}\right).
\end{equation}
We also see

\begin{eqnarray}
\hspace{-1cm} \sum_{k=0}^{q-1} e^{V_K(k+l)}
 &=&
  \sum_{k=0}^{q-1} \sum_{m\in \mathbb{Z}} 
 e^{-\frac{K}{2}\left(\frac{2\pi}{q}(k+l)-2\pi m\right)^2}
 =
 \sum_{k=0}^{q-1} \frac{1}{\sqrt{2\pi K}}
 \sum_{l'} e^{-\frac{(l')^2}{2K}+i l' \frac{2\pi(k+l)}{q}} \nonumber \\
 \hspace{-1cm} &=&
 \frac{q}{\sqrt{2\pi K}}\sum_{m\in \mathbb{Z}}
 e^{-\frac{m^2 q^2}{2K}} 
 =
 \sum_{m\in \mathbb{Z}} e^{-\frac{2\pi^2 K m^2}{q^2}}
 =
 F_{K/q^2}(0),
\end{eqnarray}
so that we have
\begin{equation}
 \label{plK}
 p_l(K) = \frac{F_K\left(\frac{2\pi k}{q}\right)}{F_{K/q^2}(0)}.
\end{equation}
Hence equations (\ref{SDnl})--(\ref{2CPnlq}) can be rewritten in terms
of $F_K(x)$.
It should be noticed that the function $F_K(x)$ can be written in 
terms of $\vartheta_3$, a Theta function:
\begin{equation}
\hspace{-2cm}
 F_K(x) 
 = 
 e^{-\frac{K}{2}x^2} \sum_{m\in \mathbb{Z}} (e^{-2\pi^2 K})^{m^2} (e^{\pi K x})^{2m}
 = 
 e^{-\frac{K}{2}x^2} \sum_{m\in \mathbb{Z}} Q^{m^2} z^{2m}
 =
e^{-\frac{K}{2}x^2} \vartheta_3(u,Q), 
\end{equation}
where $z=e^{\pi K x}=e^{ui}$ ($u=-i\pi K x$) and $Q=e^{-2\pi^2 K}$.
This expression is useful for numerical evaluations of the 
transition points on the computer with special functions software implemented.

Now we would like to discuss the plausibility of our conjectures
(\ref{SDnl}) - (\ref{2CPnlq}). 
The $q=2$ and the $q=3$ cases are equivalent to the Ising model
and the three state Potts model, respectively, both of which have been discussed 
in detail in \cite{MNN}. These are favourable facts to support our
conjectures. 
Unfortunate aspect about the $Z_q$ model is the difficultly 
of the analysis of the replicated model even for $n=1$.
Hence we do not give such evidence as we did for the $Z_2$ case.
Instead, the limit $q\gg 1$ is discussed in the next subsection.

\subsection{The limit $q \to \infty$}
\label{VInl1}

When $q$ is large and the transition point is unique,
it is reasonably expected that the transition point is of order $O(q)$.
In fact, if we set $K=\gamma q$ and suppose $q$ is large, we find a consistent
solution as follows.  The functions appearing in (\ref{plK}) behave asymptotically as
\begin{eqnarray}
 F_{K/q^2}(0) 
 &=
 \frac{q}{\sqrt{2\pi K}}\sum_{m\in \mathbb{Z}} e^{-\frac{m^2 q^2}{2 K}}
 \sim
 \frac{q}{\sqrt{2\pi K}},
 \\
 F_{\gamma q}\left(\frac{2\pi l}{q}\right) 
 &=
 \sum_{m\in \mathbb{Z}}
 e^{-\frac{\gamma}{2}(\frac{2\pi}{\sqrt{q}}l-2\pi\sqrt{q}m)^2}
 \sim
 e^{-\frac{\gamma}{2}(\frac{2\pi l}{\sqrt{q}})^2}.
 \label{ThG}
\end{eqnarray}
Substitution of these expressions into (\ref{SDnlq}) leads to
\begin{equation}
 \sqrt{2\pi \gamma}\sum_l 
 \frac{\gamma}{2\sqrt{q}}\left(\frac{2\pi l}{\sqrt{q}}\right)^2
 e^{-\frac{\gamma}{2}(\frac{2\pi l}{\sqrt{q}})^2}
 =
 \frac12 \log (2\pi \gamma).
\end{equation}
If we approximate the summation by an integral, which should be valid for
large $q$, we find
\begin{eqnarray}
 \mbox{LHS} 
 \sim
 \sqrt{2\pi \gamma} \int_{-\infty}^{\infty} dy \frac{\gamma}{2}(2\pi y)^2 
 e^{-\frac{\gamma}{2}(2\pi y)^2}
 =\frac{1}{2},
\end{eqnarray}
where we set $y=l/\sqrt{q}$. We therefore have
\begin{equation}
 \log (2\pi \gamma) = 1,
\end{equation}
implying $\gamma=e/2\pi$ with $e$ being the base of natural logarithm.
Hence, when uniqueness of the critical point is assumed,
the asymptotic location is 
\begin{equation}
\label{Kqe}
 K = \frac{q e}{2\pi}.
\end{equation}

When there are two critical points $K_1,K_2$, the above argument 
should be modified to some extent. If we assume that 
both of $K_1$ and $K_2$ are of order $O(q)$, we may set $K_i=\gamma_i q$ 
($i=1,2$) and apply the same procedure as above. 
The result is that $K_1$ and $K_2$ should be related through
\begin{equation}
\label{2CPqi}
 K_1 K_2 = \frac{q^2 e^{2}}{4\pi^2}.
\end{equation}
However, for the non-random case, it is known that 
the two transition points are not of order $O(q)$;
one is of order $O(q^2)$ and the other $O(1)$ \cite{EPS1979}.
Since this is expected to persist for the random 
case (see subsection \ref{ph_st} below), 
equation (\ref{2CPqi}) may not necessarily capture the
true asymptotics of the transition points.

\subsection{Structure of the phase diagram}
\label{ph_st}

In this subsection we discuss the phase structure of the random $Z_q$ model.
For the case of the non-random model, it was shown that, 
when $q$ is sufficiently large,
the simple Ising-like two phase picture (with ferromagnetic and paramagnetic phases)
is not possible
\cite{EPS1979}.
The authors of \cite{EPS1979} showed that the transition point determined 
by the duality in the case of only two phases is inconsistent
with a kind of Griffith's inequality. 
The discussions were based on the assumption of the existence 
of the KT transition in the continuous model $q\to\infty$.

We would like to address the issue of the full phase structure 
of the random $Z_q$ model but let us restrict our main interest
on the NL for a moment.
Basically it is expected that the phase transitions on the NL 
are of a similar nature to the non-random case.
There are three phases; disordered, KT and ferromagnetic.
This conclusion may be drawn by using the same arguments 
as for the non-random case.
There are, however, still some debates about the existence 
of the KT phase when randomness is introduced in the model \cite{Korshunov1993,Tang1996,MudryWen1999}.
Hence, in this section, we give a different argument supporting 
the three-phase picture without assuming the existence 
of the KT transition in the continuous model.
A crucial point in our arguments is the proof of existence 
and non-existence of the ferromagnetic phase in appropriate
parameter regions when $q \gg 1$.

First we show the existence of a ferromagnetic phase near the limit
$K, K_p\to\infty$ (the ground state of the non-random system)
 following 
\cite{Griffiths1964,HoriguchiMorita1982}. 
Let us first consider the non-random case.
The order parameter is bounded as
\begin{eqnarray}
 \label{Zq-orderparameter}
 \hspace{-1cm} \langle e^{2\pi i \frac{k_j}{q}} \rangle_K
 &=&
 \sum_{k=0}^{q-1} P_k e^{2\pi i \frac{k}{q}} 
 =
 P_0 + \sum_{k(\neq 0)} P_k e^{2\pi i \frac{k}{q}}
 \geq 
 P_0 -\sum_{k(\neq 0)} P_k 
 =
 1-2(1-P_0)
 \nonumber \\
 \hspace{-1cm}
 &=&
 1-2 \langle N' \rangle_K /N,
\end{eqnarray}
where $\langle \ \rangle_K$ means thermal average,
$P_k$ is the probability that the spin takes the value $k$ on 
each site, and $N'$ is the number of sites such that $k\neq 0$.
All boundary spins have $k=0$.
As in \cite{Griffiths1964,HoriguchiMorita1982}, $N'$ is 
bounded as
 \begin{eqnarray}
 N' 
 \leq 
 \sum_{b=4,6,\cdots} \left(\frac{b}{4}\right)^2 
 \sum_{j=1}^{\nu(b)}X_b^{(j)},
\end{eqnarray}
where $X_b^{(j)}$ is 1 if the $j$th border 
(separating a domain of sites from those with different $k$'s) of length $b$
occurs in a configuration and 0 otherwise.
The symbol $\nu(b)$ stands for the number of possible borders of length $b$.
The average $\langle X_b^{(j)}\rangle_K$ is bounded by, for large $q$,
 \begin{equation}
 \langle X_b^{(j)}\rangle_K 
 \leq 
 e^{-\frac{K b}{2} \left(\frac{2\pi}{q}\right)^2}. 
\end{equation}
Using 
\begin{equation}
\nu(b) \leq  4 \cdot 3^b q^b N/(3 b),
\end{equation}
where the factor $q^b$ comes from the possible number of 
boundary bonds,
we find
\begin{equation}
 \frac{\langle N' \rangle_K}{N}
 \leq 
 \frac{1}{12}\sum_{b=4,6,\cdots} 
 b 3^b q^b e^{-\frac{Kb}{2}\left(\frac{2\pi}{q}\right)^2}.
\end{equation}
Hence if we choose
$ \kappa \equiv 3 q e^{-\frac{K }{2} \left(\frac{2\pi}{q}\right)^2}$
sufficiently small, the order parameter, the left-hand side of (\ref{Zq-orderparameter}),
 is certainly positive.
 From this follows that the ferromagnetic phase exists
when $T < C_1^{(0)}/ q^2$ with $C_1^{(0)}$ being some positive constant.

Following \cite{HoriguchiMorita1982}, we can generalize 
the above argument to the 
random case. At least for $K$ and $K_p$ very large,
it is possible to prove the existence of a ferromagnetic phase.
Let us take a border of spin configurations such that $X_b^{(j)}\neq 0$ 
and notice that the thermal average $ \langle X_b^{(j)} \rangle_K $
is written as
\begin{equation}
\label{Xss}
 \langle X_b^{(j)} \rangle_K 
 =
 \frac{ \sum_{\{k\}}'\prod_{\langle ij \rangle } 
        e^{V_K(k_{i}-k_{j}-l_{ij})}}
      { \sum_{\{k\}}\prod_{\langle ij \rangle } 
        e^{V_K(k_{i}-k_{j}-l_{ij})}},
\end{equation}
where the sum in the numerator is taken over states in 
which the $j$th border of length $b$ occurs.
Here $l_{ij}$ is the quenched randomness and occurs with 
probability $r^{l_{i,j}^2}$ ($r=e^{-2\pi^2 K_p/q^2}$) 
approximately at each bond. 
Take $b=2n$ ($n$ to be distinguished from the replica number)
and consider the case where the number 
of such bonds with $l_{i,j}\neq 0$ is $m$. 
When $m=n,n+1,\cdots,2n$, 
one restricts the summation in the denominator of (\ref{Xss}) to 
configurations which appear in the numerator and obtains 
a trivial upper bound $\langle X_{2n}^{(j)} \rangle \leq 1$.
On the other hand, when $m=0,1,\cdots,n-1$, one finds an upper bound 
$\langle X_{2n}^{(j)} \rangle \leq \lambda^{2n-m-\sum l_{i,j}^2 }$ 
with $\lambda=e^{-2\pi^2 K/q^2}$, where the summation in the 
exponent is over the bonds with $\l_{i,j}\neq 0$. 
This is obtained by restricting the
summation in the denominator in (\ref{Xss}) to 
configurations with all $k_i =0$.
Hence the configurational average 
$[\langle X_{2n}^{(j)} \rangle_K]_{\rm av}$ is bounded as
\begin{equation} 
\hspace{-5mm} [\langle X_{2n}^{(j)}\rangle_K]_{{\rm av}}
\leq
\sum_{m=0}^{n-1} \left[\begin{array}{c}2n\\m\end{array}\right] 
                 \sum_{l_{i,j}} r^{\sum l_{i,j}^2}
                 \lambda^{2n-m-\sum l_{i,j}^2}
+
\sum_{m=n}^{2n} \left[\begin{array}{c}2n\\m\end{array}\right] q^m r^m.
\end{equation}
When $K\leq K_p$, one has $r/\lambda \leq  1$ so that this can be 
replaced by
\begin{equation} 
\hspace{-5mm} [\langle X_{2n}^{(j)}\rangle_K]_{{\rm av}}
\leq
\sum_{m=0}^{n-1} \left[\begin{array}{c}2n\\m\end{array}\right] 
                 q^m r^m \lambda^{2n-2m} 
+
\sum_{m=n}^{2n} \left[\begin{array}{c}2n\\m\end{array}\right] q^m r^m.
\end{equation}
Then, if we choose $K_p/q^2 \geq K/q^2\gg 1$, 
by almost the same argument as in the non-random case, 
$\langle N' \rangle_K /N$ can be made sufficiently small 
so that the order parameter does not vanish.
In particular, there exists a positive constant $C_1$ such that 
the order is nonzero when $T < C_1 / q^2$ on the NL.
The ferromagnetic phase is also expected when $K_p/q^2 \geq K/q^2\gg 1$
but we need a little tighter estimation to prove it.

On the other hand, to show non-existence of ferromagnetic phase at high temperature,
we use the arguments in \cite{HoriguchiMorita1981}.
We show below
\begin{equation}
\label{order_pure}
 \left|\left[\langle e^{2\pi i\frac{k_j}{q}} \rangle_K \right]_{\rm av}\right|
 \leq
 \langle e^{2\pi i k_j/q} \rangle_{K_p,\rm{nonrandom}}
 \langle e^{2\pi i k_j/q} \rangle_{K  ,\rm{nonrandom}},
\end{equation}
where the average on the right-hand side is taken for the non-random model 
at inverse temperatures $K_p$ and $K$, respectively.
To verify this inequality, we first notice
\begin{eqnarray}
 \hspace{-2.5cm}
 \left[\langle e^{2\pi i\frac{k_j}{q}} \rangle_K \right]_{\rm av}
 \hspace{-3mm}&=&
 \left[\frac{\Tr_{\{k\}}e^{2\pi ik_j/q}
             \prod_{\langle j_1,j_2 \rangle} 
             e^{V_K(k_{j_1}-k_{j_2}-l_{j_1,j_2})}}
            {\Tr_{\{k\}} \prod_{\langle j_1,j_2 \rangle} 
             e^{V_K(k_{j_1}-k_{j_2}-l_{j_1,j_2})}} \right]_{\rm av} \nonumber \\
 \hspace{-2.5cm}
 \hspace{-3mm}&=&
 \sum_{\{l\}} \left( \prod_{\langle j_1,j_2 \rangle} 
 \frac{e^{V_{K_p}(l_{j_1,j_2})}}{\sum_{l_1,l_2}e^{V_{K_p}(l_{j_1,j_2})}}\right)
 \frac{\Tr_{\{k\}}e^{2\pi ik_j/q} \prod_{\langle j_1,j_2 \rangle} 
             e^{V_K(k_{j_1}-k_{j_2}-l_{j_1,j_2})}}
            {\Tr_{\{k\}} \prod_{\langle j_1,j_2 \rangle} 
             e^{V_K(k_{j_1}-k_{j_2}-l_{j_1,j_2})}}.
\end{eqnarray}
Applying the gauge transformation,
\begin{eqnarray}
 k_j         & \to& k_j -\kappa_j, \\
 l_{j_1,j_2} & \to& l_{j_1,j_2} - \kappa_{j_1}+ \kappa_{j_2},
\end{eqnarray}
we find
\begin{equation}
\label{Pl}
 \left[\langle e^{2\pi i\frac{k_j}{q}} \rangle_K \right]_{\rm av}
 =
 \sum_{\{l\}}  P_{\{l\}} \langle e^{2\pi i\frac{\kappa_j}{q}} \rangle_{K_p}
                         \langle e^{2\pi i\frac{k_j}{q}} \rangle_K,
\end{equation}
where $P_{\{l\}}$ denotes a certain probability distribution.
If we apply a very reasonable inequality that the value of the order parameter of the random
system is smaller than its non-random counterpart with the same inverse temperature
\begin{equation}
 |\langle e^{2\pi i \kappa_j/q} \rangle_{K_p}|
 |\langle e^{2\pi i  k_j/q} \rangle_{K}|
 \leq
 \langle e^{2\pi i \kappa_j/q} \rangle_{K_p,\rm{nonrandom}}
 \langle e^{2\pi i  k_j/q} \rangle_{K  ,\rm{nonrandom}}
\label{ran_pure}
\end{equation}
to each term in (\ref{Pl}), we arrive at (\ref{order_pure}).
Unfortunately we have not succeeded in proving the inequality 
(\ref{ran_pure}) mathematically except for the $q=2$ case, 
which was already proved in \cite{HoriguchiMorita1979}.
However, even in the absence of a formal mathematical proof,
equation (\ref{ran_pure}) should certainly be valid.
Now, if we let $K_1$ denote the phase transition point of 
the non-random model at which the order vanishes, we see that the 
ferromagnetic phase cannot exist when $K<K_1$ or $K_p < K_1$.
In fact $K_1$ is known to be of order $O(q^2)$ \cite{EPS1979} and we 
conclude that the ferromagnetic phase cannot exist on the NL
when $T> C_2/q^2$ with $C_2$ some positive constant.

Notice that the above arguments imply the existence of (at least one)
transition point(s) between ferromagnetic and non-ferromagnetic phases.
Now if there is a unique transition point,
it is of order $O(1/q)$ [cf (\ref{Kqe})]. 
But this contradicts with the fact that the order vanishes
when $T> C_2/q^2$. Therefore we have shown that the simple Ising-like
two phase picture is not possible when $q$ is 
large enough.
This conclusion strongly suggests that there are three phases 
in the full phase diagram. A schematic phase diagram expected from the 
above arguments is given in figure 3.
\begin{figure}
\begin{center}
  \includegraphics[width=6cm]{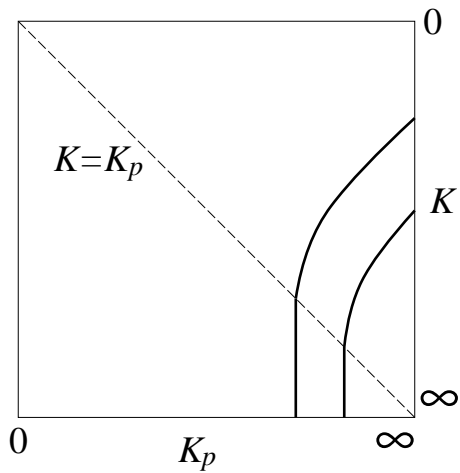}
\label{phasezq}
\caption{\small Phase diagram of the random $Z_q$ model. Solid lines 
show expected phase boundaries.}
\end{center}
\end{figure}

As a remark,
for $q$ not very large, the situation may be more subtle.
Numerical simulation results suggest 
that something peculiar may be happening for these models
(e.g. NER analysis in \cite{Ozeki}).
Further careful analyses are necessary on this point.

\subsection{Maximally Random Case}
\label{m0}
Let us consider the maximally random case ($K_p=0$),
in which the possibility of a finite temperature phase transition has been 
discussed in the literature \cite{HolmeKimMinnhagen2003}. 
Here we give an argument against such a possibility using duality.

As in the case of the Ising model \cite{MNN}, the $n$-replicated 
system with $K_p=0$ is related to the $(n-1)$-replicated system
on the NL. More precisely, it is easy to prove
\begin{equation}
 \label{ZnZn-1}
 {\cal Z}_n(K,K_p=0) \propto {\cal Z}_{n-1}(K,K_p=K),
\end{equation}
where ${\cal Z}_n(K,K_p)$ stands for the averaged partition function (\ref{Zisotropy}).
Now let us assume the applicability of this relation down to the 
$n\to 0$ limit and discuss the phase transitions for 
the quenched model with $K_p=0$. Note that the $n\to 0$ limit
for $K_p=0$ is equivalent to taking the $n\to -1$ limit 
on the NL according to (\ref{ZnZn-1}).

If we assume that there is a unique transition point,
it is determined by the relation (\ref{SDnl}), which reads
\begin{equation}
\label{Km}
 \frac{\sum_{l=0}^{q-1}
 e^{nV_K(l)}}{\left(\sum_{l=0}^{q-1}e^{V_K(l)}\right)^n} = q^{-n+1}.
\end{equation}
Let us take the $n\to 0$ limit. If we set $K = c/n$, we see
\begin{equation}
 \left(\sum_{l=0}^{q-1}e^{V_K(l)}\right)^n  \sim e^{V_c(0)},
\end{equation}
so that (\ref{Km}) becomes
\begin{equation}
 \frac{\sum_{l=0}^{q-1} e^{V_c(l)}}{e^{V_c(0)}} = q.
\end{equation}
This is nothing but a condition to determine the 
transition point for the non-random system when there is a 
unique transition point.
Accordingly we know that $c=q/(2\pi)$ by (\ref{K1K2Zq}) with $K_1=K_2$ and hence 
$K=c/n \to \infty$ as $n\to 0$. 
This means that the phase transition, if it is unique, occurs at $T=0$.
The same discussions can be applied to the case where 
there are two transition points. In any case the conclusion is that
the phase transitions, if there exist, occur only at $T=0$.

We should remember that our discussions were based on the 
subtle assumption about the applicability of the relation as $n\to 0$
($n\to -1$ on the NL) and the assumption that (\ref{SDnl}) or (\ref{2CPnl}) gives
the criticality condition.
Hence our conclusion has not been completely rigorously derived.
It would be interesting to clarify this point with 
other methods or arguments.


\section{Conclusion and discussion}

In this paper, we have generalized the duality argument 
combined with the replica method, which was originally applied
to the $2d$ random bond Ising model on the square lattice, to a variety of 
random spin systems. Our main results are the conjectures 
on the transition points of systems with quenched randomness.

First we considered the random $Z_2$ models.
We have given the conjecture (\ref{dualcriticalpoint}) 
about the relation between the multicritical points of 
two models which are mutually dual. 
Besides exact computations for the replicated system with $n=1, 2$ and $\infty$,
numerical simulations support our conjecture for the $2d$ $\pm J$ random bond Ising
model on the 
triangular/hexagonal lattices, the $3d$ random Ising/gauge models on the
cubic lattice and the $2d$ anisotropic random bond Ising model on the 
square lattice.
We think it remarkable that a single theoretical framework makes it possible
to derive a series of predictions to be compared favourably with many
independent numerical simulations.

Next we treated the random $Z_q$ model. We have shown that  
there are at least two transition points for sufficiently large $q$. 
The most probable scenario is that there are three phases in the phase diagram, 
paramagnetic, KT-like and ferromagnetic phases.
Our arguments, however, do not assume the existence or some specific
properties of the KT phase.
By applying the duality argument to this model,
we have also given the conjectured relation (\ref{2CPnlq}) 
between the two transition points.

An interesting question is why we restrict ourselves to the NL.
The fixed-point condition of the leading Boltzmann factor $x_0(K, p)=x_0^*(K, p)$
relates $p$ and $K$ and may give the whole shape of the phase boundary.
This is indeed the case for $n=1$ as well as for $n=2$ above the multicritical point.
We nevertheless have restricted ourselves to the NL in this paper because
it is difficult to directly verify the ansatz on the whole part of
the phase diagram numerically for many systems with quenched randomness.
It should also be kept in mind that the multicritical point is the place
where two completely different types of symmetries, invariance under
duality and gauge transformation, meet under the present conjecture.
This implies that the multicritical point has clearly distinguished
symmetry features, which allows us to discuss this point on a different
basis from other points of phase transition.

We believe that sufficient analytical and numerical 
evidence has been accumulated to support the validity of our conjecture. 
It is an interesting future problem to provide a mathematically rigorous proof.

 \ack
 The authors would like to thank 
 Y Ozeki and T Nakamura for their suggestions and advice 
 on numerical calculations and T Hamasaki for his useful comments.
 This work was supported by the Grant-in-Aid for Scientific Research
 on Priority Area
 ``Statistical-Mechanical Approach to Probabilistic Information
 Processing'' and the 21st Century COE Program at Tokyo Institute of
 Technology ``Nanometer-Scale Quantum Physics''.  
 TS is also supported by the Grant-in-Aid for Young 
 Scientists (B), the Ministry of Education, Culture, Sports, Science and 
 Technology, Japan.


\section*{Appendix}

In this appendix we calculate the prefactor in the duality relation 
of the partition function (\ref{o_d}) in section 2.
Here we follow the derivation of the duality relation
by Wu and Wang \cite{WuWang}.

The partition function of the system described by the Hamiltonian
(\ref{generalspin}) in section 2 is
\begin{equation}
{\cal Z} = \sum_{\{\xi_{\scriptscriptstyle \xv}=0,1\}}  \prod_{C} U(\xi_{C}), 
\label{appZ}
\hspace{0.6cm}
\end{equation}
where
\begin{equation}
 U(\xi_{C}) =  \prod_{{\xv} \in \partial C}
 \exp \left( K  \xi_{\xv} \right).
\end{equation}
$U(\xi_{C})$ is the Boltzmann factor for an element $C$ with dimension $r$.
For example, the usual two-body interaction on a bond has $r=1$ and the lattice
gauge theory has $r=2$.
$\xi_{\xv}$ is a modulo-2 spin variable which takes 0 or 1, and
$\xi_{C}$ is defined by $\xi_{C}=\sum_{{\xv} \in \partial C} \xi_{\xv}$
modulo 2.

Let us define the dual Boltzmann factor for the dual element $C^{*}$ 
by the binary Fourier transformation,
\begin{equation}
U^{*} (\lambda_{C^{*}}) = 2^{-\frac{1}{2}}\sum_{\xi_C = 0,1}
 \exp (2 \pi i \xi_C \lambda_{C^{*}}) U (\xi_{C}),
\end{equation} 
or conversely,
\begin{eqnarray}
U (\xi_{C}) &=& 2^{-\frac{1}{2}}\sum_{\lambda_{C^{*}} = 0,1}
 \exp (2 \pi i \lambda_{C^{*}} \xi_C) U^{*} (\lambda_{C^{*}}) \nonumber \\
&=& 2^{-\frac{1}{2}}\sum_{\lambda_{C^{*}} = 0,1}
 \exp \left( 2 \pi i  \lambda_{C^{*}} \sum_{{\xv} \in \partial C} \xi_{\xv}
 \right)
 U^{*} (\lambda_{C^{*}}) \nonumber \\
&=& \sum_{\lambda_{C^{*}} = 0,1}
 \left( \prod_{{\xv} \in \partial C} T(\xi_{\xv}, \lambda_{C^{*}}) \right)
 U^{*} (\lambda_{C^{*}}),
\label{appFourier}
\end{eqnarray} 
where
\begin{equation}
T(\xi_{\xv}, \lambda_{C^{*}}) \equiv 2^{-\frac{1}{2 B_C}}
 \exp (2 \pi i \lambda_{C^{*}} \xi_{\xv}),
\end{equation}
and $B_C$ is the number of ($r-1$)-dimensional elements on the boundary of $C$.

Inserting (\ref{appFourier}) into (\ref{appZ})
 and taking the sum over $\xi_{\xv}$, we
obtain a modulo-2 Kronecker delta for each $\xi_{\xv}$,
\begin{eqnarray}
& & \sum_{\xi_{\xv}=0,1} \ \prod_{C^{*}: {\xv} \in \partial C} 
T(\xi_{\xv}, \lambda_{C^{*}}) \nonumber \\
&=& \sum_{\xi_{\xv}=0,1} \ \prod_{C^{*}: {\xv} \in \partial C} 
2^{-\frac{1}{2 B_C}}
 \exp (2 \pi i \lambda_{C^{*}} \xi_{\xv}) \nonumber \\
&=& 2^{1-\sum_{C: {\xv} \in \partial C} \frac{1}{2 B_C}}
\delta_{\rm mod 2} \left( \sum_{C^{*}: {\xv} \in \partial C} 
\lambda_{C^{*}} \right). \label{appFT2}
\end{eqnarray}
It is useful to define the following symbol of a constrained sum,
which stems from the Kronecker deltas in (\ref{appFT2}),
\begin{equation}
\label{appsum}
{\sum_{\{\lambda_{C^{*}}=0,1\}}}^{\!\!\!\!\!\! '} 
\equiv \sum_{\{\lambda_{C^{*}}=0,1\}} \prod_{\xv} \delta_{\rm mod 2} \left( \sum_{C^{*}: {\xv} \in \partial C} \lambda_{C^{*}} \right). 
\end{equation}
Using this, (\ref{appZ}) becomes
\begin{equation}
{\cal Z} = \left( \prod_{\xv} 2^{1-\sum_{C: {\xv} \in \partial C} \frac{1}{2 B_C}} \right) \left( { \sum_{ \{ \lambda_{C^{*}}=0,1 \} }}^{\!\!\!\!\!\!\! '} \  \prod_{C}  U^{*} (\lambda_{C^{*}}) \right).
\end{equation}
The prefactor can be simplified as 
\begin{eqnarray}
& & \prod_{\xv} 2^{1-\sum_{C: {\xv} \in \partial C} \frac{1}{2 B_C}}
\nonumber \\
&=& 2^{N_{r-1} - \frac{1}{2} \sum_{\xv} \sum_{C: {\xv} \in \partial C} \frac{1}{B_C}}
\nonumber \\
&=& 2^{N_{r-1} - \frac{N_{r}}{2}},
\end{eqnarray}
where the number of $m$-dimensional elements is denoted by $N_{m}$.

The final result is
\begin{equation}
\label{appZorig}
{\cal Z} = 2^{N_{r-1} - \frac{N_{r}}{2}} {\sum_{\{\lambda_{C^{*}}=0,1\}}}^{\!\!\!\!\!\!\! '} \ \prod_{C^{*}}  U^{*} (\lambda_{C^{*}}).
\end{equation}
Here we have replaced the product over $C$ with $C^{*}$, 
which are identical operations.

The partition function (\ref{appZ}) can also
be rewritten using the Boltzmann factors for the element $C$,
\begin{eqnarray}
\label{appZdual1}
{\cal Z} &=& \sum_{\{\xi_{\xv}=0,1\}}  \prod_{C} U(\xi_{C}) \\
\label{appZdual2}
 &=& 2^{{\cal N}_{g}} {\sum_{\{\xi_{C}=0,1\}}}^{\!\!\!\!\!\! '}
 \ \prod_{C} U(\xi_{C}).
\end{eqnarray}
Note that $U(\xi_{C})$ in (\ref{appZdual1}) is a function of $\xi_{\xv}$
as in (1), but it is not in (\ref{appZdual2}): It is
 redefined on each element $C$ as a function of $\xi_{C}$ itself.  
The usual sum over $\xi$ in (\ref{appZdual1}) can be written
 as in (\ref{appZdual2})
using the definition (\ref{appsum})
if we consider a
correct mapping from the configuration space defined by $\xi_{\xv}$
 onto the one by $\xi_{C}$.
The prefactor $2^{{\cal N}_g}$ in (\ref{appZdual2}) is the degree of 
the ground-state degeneracy.
The mapping from the $\xi_{\xv}$ space to the $\xi_{C}$ space is not 
one-to-one. In (\ref{appZdual2}) the ground state configuration is
 $\xi_{C}=0$ for all $C$ for ferromagnetic interactions and is unique,
 while in (\ref{appZdual1}) 
the ground state is degenerate. Thus the mapping is $2^{{\cal N}_g}$ to 1.
 For example, $2^{{\cal N}_g}=2$ for $r=1$ because all-up and all-down
 states are degenerate. For $r=2$,
 $2^{{\cal N}_g}$ is dependent on the number of sites because
 the model has $Z_2$ gauge symmetry in this case.
Wegner calculated this 
degree of degeneracy in \cite{Wegner} under general conditions
 and the result is
\begin{eqnarray} 
\label{appdeg}
{\cal N}_{g} &=& \sum_{m=0}^{r-2} (-1)^{r-m} N_m + t_{1}
 \hspace{1.5cm} {\rm for}\
 r\geq2, \nonumber \\
      &=& 1 \hspace{4.9cm} {\rm for}\ r=1,
\end{eqnarray}
where $t_1$ is a constant which depends on the topology of the lattice
 (or the boundary condition) and not on the number of elements.
If we use the generalized Euler's relation for the number of lattice
elements \cite{Wegner},
\begin{equation}
\label{appEuler}
\sum_{m=0}^{d} (-1)^{m} N_m = t_2,
\end{equation}
where $t_2$ is also a constant dependent on the topology, ${\cal N}_g$ becomes 
\begin{equation}
\label{appNg}
{\cal N}_{g} = \sum_{m=r-1}^{d} (-1)^{m-(r-1)} N_m + t,
\end{equation}
with $t$ being another constant determined by $t_1, t_2$ and $r$.

We can derive the factor (\ref{appNg}) intuitively from the
difference of the numbers of spin configurations between the expressions
(\ref{appZdual1}) and (\ref{appZdual2}).  
In (\ref{appZdual1}) the number of independent $\xi_{\xv}$'s is
$N_{r-1}$. In (\ref{appZdual2}) the number of independent $\xi_{C}$'s
may appear to be $N_{r}$, but 
we must take the number of constraints in (\ref{appZdual2}) into account, 
which is $N_{r+1}$ and should be subtracted from $N_r$.
 However these constraints are redundant or not independent
 of each other because $r+1$ dimensional element is always 
 on the boundary of $r+2$ dimensional element. 
 Therefore we must subtract
 $N_{r+2}$ from the number of constraints $N_{r+1}$. However
 the number $N_{r+2}$ is also redundant again and 
 $N_{r+3}$ must be subtracted from it, which is similar 
 for higher dimensional elements.
 Then ${\cal N}_{g}$ is calculated as
 \begin{eqnarray}
 {\cal N}_g &=&  N_{r-1} - 
 \left\{ N_{r} - (N_{r+1} - (N_{r+2} - (N_{r+3} \ldots \right\}
 \nonumber \\
 &=& \sum_{m=r-1}^{d} (-1)^{m-(r-1)} N_m + t,
 \end{eqnarray}
 which is equivalent to (\ref{appNg}).
 The constant $t$ will be the same as in (\ref{appNg}).
 
 Using (\ref{appZorig}), (\ref{appZdual2}) and (\ref{appNg}),
 we obtain the duality relation between the original and the dual
 partition functions,
\begin{eqnarray}
\label{appZdualrelation}
\hspace{-5mm} &&  2^{{\cal N}_g}
 {\sum_{\{\xi_{C}=0,1\}}}^{\!\!\!\!\!\!\! '} \ \prod_{C}  U (\xi_{C})
\hspace{5mm} \left( = {\cal Z_{\rm orig}}\{ u_1(K),u_{-1}(K) \} \right)
 \nonumber \\
\hspace{-5mm} &=&  2^{{\cal N}_{g}^{*} + a}
 {\sum_{\{\lambda_{C^{*}}=0,1\}}}^{\!\!\!\!\!\!\! '} 
\ \prod_{C^{*}}  U^{*} (\lambda_{C^{*}})
\hspace{5mm} \left( = 2^{a} {\cal Z_{\rm dual}}
\{ u_1^{*}(K),u_{-1}^{*}(K) \} \right),
\end{eqnarray}
where
\begin{equation}
{\cal N}_{g}^{*} = \sum_{m=0}^{r+1} (-1)^{m-(r+1)} N_m + t^{*},
\end{equation}
and $t^{*}$ is also a constant which depends on the topology.
$a$ is defined by
\begin{equation}
 a = - {\cal N}_{g}^{*} + N_{r-1} - \frac{N_{r}}{2}.
\end{equation}
Remember that we use the modulo-2 spin variables here, 
while spins take the value $\pm 1$ in the main text.

Now let us consider the self-dual case. In this case $d$ must be even
and $r=d/2$. Furthermore the number of $m$ dimensional elements satisfies
\begin{equation}
\label{appNdual}
N_{m} = N_{d-m}.
\end{equation}
Inserting this into the Euler's relation (\ref{appEuler}), we obtain
\begin{equation}
 \hspace{-1cm}
  \sum_{m=0}^{d/2-1} (-1)^{m} N_m + (-1)^{d/2} \frac{N_{d/2}}{2} =
  \sum_{m=d/2+1}^{d} (-1)^{m} N_m + (-1)^{d/2} \frac{N_{d/2}}{2} =
  \frac{t_{2}}{2}.
\end{equation}
In addition, ${\cal N}_{g}$ and ${\cal N}_{g}^{*}$ become
\begin{eqnarray}
\hspace{-1cm} {\cal N}_{g} &=& \sum_{m=d/2-1}^{d} (-1)^{m-(d/2-1)} N_m + t = 
 N_{d/2-1} - \frac{N_{d/2}}{2} + (-1)^{d/2-1}  \frac{t_2}{2} + t,  
\nonumber \\
\hspace{-1cm} {\cal N}_{g}^{*} &=& \sum_{m=0}^{d/2+1}
 (-1)^{m-(d/2+1)} N_m + t^{*}
 = N_{d/2+1} - \frac{N_{d/2}}{2} + (-1)^{d/2+1} \frac{t_2}{2} +t^{*}
 \nonumber \\
\hspace{-1cm} &=&  N_{d/2-1} - \frac{N_{d/2}}{2} + (-1)^{d/2-1} \frac{t_2}{2} +t^{*} = {\cal N}_{g} + t^{*} - t. 
\end{eqnarray}
Hence,
\begin{equation}
 a = (-1)^{d/2} \frac{t_2}{2} - t^{*}.
\end{equation}
Therefore
$a$ is just a trivial constant in the self-dual case
and negligible in the thermodynamic limit.
The factors $2^{{\cal N}_{g}}$ and $2^{{\cal N}_{g}^{*}}$
 in (\ref{appZdualrelation}),
 which differ only by a trivial constant, 
do not concern when we derive the transition point from
the relation $u_{\pm 1}(K_c)=u^{*}_{\pm 1}(K_c)$.


\end{document}